\documentstyle[preprint,aps]{revtex}
\begin{document}
\draft
\preprint{}
\title{Axially Symmetric Solutions for SU(2) Yang-Mills Theory}
\author{D. Singleton}
\address{Department of Physics, University of Virginia,
Charlottesville, VA 22901}
\date{\today}
\maketitle
\begin{abstract}
By casting the Yang-Mills-Higgs equations of an SU(2) theory
in the form of the Ernst equations of general relativity, it
is shown how the known exact solutions of general relativity can be
used to give similiar solutions for Yang-Mills theory. Thus
all the known exact solutions of general relativity with axial
symmetry ({\it e.g.} the Kerr metric, the Tomimatsu-Sato metric)
have Yang-Mills equivalents. In this paper we only examine in detail
the Kerr-like solution. It will be seen that this
solution has surfaces where the gauge and scalar fields become
infinite, which correspond to the infinite redshift surfaces of the
normal Kerr solution. It is speculated that this feature may be
connected with the confinement mechanism since any particle which
carries an SU(2) color charge would tend to become trapped once it
passes these surfaces. Unlike the Kerr solution, our solution
apparently does not have any intrinsic angular momentum, but rather
appears to give the non-Abelian field configuration associated with
concentric shells of color charge.
\end{abstract}
\newpage
\narrowtext
\section{Introduction}

Recently, using the long known connection between general relativity
and Yang-Mills theories \cite{utiyama}, we found exact Schwarzschild-like
solutions for Yang-Mills theories \cite{sing1} \cite{sing2}, which were
conjectured to have a possible connection with the confinement problem.
Our results were conceptually similiar to several other recent
papers \cite{lunev} which discussed  related solutions.
A natural question which arises from this is
if there are other exact solutions of general relativity
which have corresponding exact Yang-Mills solutions. Of particular
interest is the Kerr solution which has an intrinsic angular momentum.
In this paper we show that there are such solutions by considering an
SU(2) gauge theory coupled to a scalar field in the adjoint representation.
The solutions are found by first transforming
the Yang-Mills field equations into the Ernst equations
\cite{ernst} of general relativity, and then applying the form
of the general relativistic solution in terms of the gauge and
scalar fields. Even though we specialize in this paper to the Kerr-like
solution, it is in principle possible to use this same procedure to
map over any axially symmetric solutions to Einstein's equations
into an equivalent  Yang-Mills solution. However, as we will show,
even the Kerr-like solution has a very complex structure which
makes it difficult to deal with. In addition it may be
possible to reverse the above method and use known exact
solution of Yang-Mills theory ({\it e.g.} the BPS dyon solution
\cite{bogo} \cite{prasad} and multimonopole solutions \cite{forg1})
to write down undiscovered solutions to Einstein's
equations. One of the most interesting features of our
previous Schwarzschild-like solution was the existence of a spherical
shell surrounding the origin, on which the gauge and scalar
fields became infinite, implying the presence of a spherical
distribution of color charge. The Yang-Mills Kerr
solution, in contrast, has two concentric shells of SU(2)
charge (these shells are the equivalent of the infinite
redshift surfaces of the normal Kerr solution). These infinite field
surfaces might yield a possible mechanism for confinement, since
just as general relativistic black holes permanently trap any object
that carries gravitational ``charge'' ({\it i.e.} mass-energy), and
which moves inside the event horizon, so the Yang-Mills version
of these solutions may confine any particle which carries the gauge
charge and crosses the color event horizons. Actually, what we
call the color event horizon of our solution, corresponds to the
infinite redshift surfaces rather than the true event horizons of the
normal Kerr metric. The reason for calling these surfaces the
color event horizons is that the physical quantities of our
theory (the gauge and scalar fields) develop real singularities
on these surfaces. These infinite values of the fields imply,
at least classically, that a particle carrying a color charge would
either be strongly repelled or strongly attracted by these surfaces.
For general relativity the corresponding singular surfaces are
coordinate singularities which arise because of the particular
coordinates that one chooses. This can best of seen for the normal
Schwarzschild solution where, by transforming to Kruskal coordinates,
one can eliminate the singularity in the metric at the Schwarzschild
radius. In addition to this difference in the nature of the
singularities, it is shown that while the regular Kerr
solution has some angular momentum, our Yang-Mills version does not.
These differences arise because the symmetries of general
relativity are space-time symmetries, while the Yang-Mills
symmetries are internal Lie symmetries.

\section{The Kerr-Like Solution}

The theory which we consider is an SU(2) gauge field which is coupled
to a scalar field in the adjoint representation, which has no
self-interaction or mass terms. The Lagrangian for this theory is
\begin{equation}
\label{lagrange}
{\cal L} = -{1 \over 4} F^{\mu \nu a} F_{\mu \nu} ^a + {1 \over 2}
D_{\mu} (\phi ^a) D^{\mu} (\phi ^a)
\end{equation}
where
\begin{equation}
F_{\mu \nu} ^a = \partial _{\mu} W_{\nu} ^a - \partial _{\nu} W_{\mu} ^a
+ g \epsilon^{abc} W_{\mu} ^b W_{\nu} ^c
\end{equation}
and
\begin{equation}
D_{\mu} \phi ^a = \partial _{\mu} \phi ^a + g \epsilon ^{abc} W_{\mu} ^b
\phi ^c
\end{equation}
To obtain the Bogomolny field equations associated with this theory one
finds the field configuration which produces an extremum in the
canonical Hamiltonian
\begin{equation}
\label{ham}
{\cal H} = \int d^3 x \left[ {1 \over 4} F_{ij} ^a F^{aij} - {1 \over 2}
F_{0i}^a F^{a0i} + {1 \over 2} D_i \phi ^a D^i \phi ^a - {1 \over 2}
D_0 \phi ^a D^0 \phi ^a \right]
\end{equation}
We now introduce an explicit scale factor for the scalar field
({\it i.e.} $\phi ^a \rightarrow A \phi ^a$) so that we can study the
special case with no scalar field by simply taking $A=0$. Next we
require that all the fields are time independent and that the
time components of the gauge fields are proportional to the scalar
fields ({\it i.e.} $W_0 ^a = C \phi ^a$, where $\phi ^a$ is the
rescaled field). Thus $W_0 ^a$ acts like an additional Higgs field
except that its kinetic term appears with the opposite sign in
Eq. (\ref{ham}). Using all these conditions and the antisymmetry
of $\epsilon ^{abc}$ we find that $D_0 \phi ^a = 0$ and $F_{0i} ^a
= C (D_i \phi ^a)$, so that the Hamiltonian becomes
\begin{eqnarray}
\label{ham1}
{\cal H} = \int d^3 x \Bigg[&& {1 \over 4} \big( F_{ij} ^a -
\epsilon_{ijk} \sqrt{A^2 - C^2} D^k \phi ^a \big) \big( F^{aij}
- \epsilon _{ijl} \sqrt{A^2 - C^2} D^l \phi ^a \big) \nonumber \\
&& +{1 \over 2} \epsilon _{ijk} \sqrt{A^2 - C^2} F^{aij} D^k \phi ^a
\Bigg]
\end{eqnarray}
Using the relationship $\epsilon_{ijk} F^{aij} D^k \phi ^a = \partial ^i
( \epsilon_{ijk} F^{ajk} \phi ^a )$ the last term in Eq. (\ref{ham1})
can be turned into a surface intergral, which in the usual development
\cite{bogo} is proportional to the magnetic charge carried by the
fields due to the topology of the Higgs field at infinity \cite{arafune}.
The lower limit of the above Hamiltonian can be found be requiring
\begin{equation}
\label{bogo}
F_{ij}^a = \sqrt{A^2 - C^2} \; \epsilon_{ijk} D^k \phi ^a
\end{equation}
which are the usual Bogomolny field equations, with the presense
of the scalar and time component of the gauge fields  explicitly
displayed through the constants $A$ and $C$. This will make it
easier to examine several special cases later on.

Several exact solutions to the field equations of this
theory have been found which possess spherical symmetry : the
Bogomolny-Prasad-Sommerfield dyon solution \cite{bogo} \cite{prasad},
and more recently a Schwarzschild-like solution \cite{sing1}. In
this paper we are looking for an axially symmetric solution. Several
authors \cite{manton} \cite{forg} have already given
an axially symmetric ansatz for the gauge and scalar fields
\begin{eqnarray}
\label{ansatz}
\Phi ^a = (0, \phi _1 , \phi _2) \hspace{1.0in} W_{\phi} ^a =
-(0, \eta _1, \eta _2) \nonumber \\
W_z ^a - (W_1, 0, 0) \hspace{1.0in}  W_{\rho} ^a = -(W_2, 0, 0)
\end{eqnarray}
where $\varphi, z, \rho$ are the usual polar coordinates and
$\phi _i$, $\eta _i$, $W_i$ are functions of $\rho , z$ only.
With this ansatz the the Bogomolny equations of Eq. (\ref{bogo})
become \cite{manton}
\begin{eqnarray}
\label{bogoax}
\rho \sqrt{A^2 - C^2} (\partial _{\rho} \phi _1 - W_2 \phi _2) &=&
- (\partial _z \eta _1 - W_1 \eta _2) \nonumber \\
\rho \sqrt{A^2 - C^2} (\partial _{\rho} \phi _2 + W_2 \phi _1) &=&
- (\partial _z \eta _2 + W_1 \eta _1) \nonumber \\
\rho (\partial _{\rho} W_1 - \partial _z W_2) &=&  \sqrt{A^2 - C^2}
(\phi _1 \eta _2 - \phi _2 \eta _1) \nonumber \\
\rho \sqrt{A^2 - C^2} (\partial _z \phi _1 - W_1 \phi _2) &=&
(\partial _{\rho} \eta _1 - W_2 \eta _2) \nonumber \\
\rho \sqrt{A^2 - C^2} (\partial _z \phi _2 + W_1 \phi _1) &=&
(\partial _{\rho} \eta _2 + W_2 \eta _1)
\end{eqnarray}
If one defines two new functions,
$f(\rho , z)$ and $\psi (\rho , z)$, such that the fields, $\eta _i ,
\phi _i$ and $W_i$ are written as
\begin{eqnarray}
\label{gfields}
\phi _1 &=& - W_1 = {1 \over f} { \partial \psi \over \partial z}
\nonumber \\
\eta _1 &=& \rho W_2 = - {\rho \over f} {\partial \psi \over
\partial \rho} \nonumber \\
\phi_2 &=& -{1 \over f} {\partial f \over \partial z} \nonumber \\
\eta _2 &=& {\rho \over f} {\partial f \over \partial \rho}
\end{eqnarray}
then the Bogomolny equations of Eq. (\ref{bogoax}) become
\cite{forg1}
\begin{equation}
\label{ernst}
Re(\varepsilon ) \nabla ^2 \varepsilon =  \nabla \varepsilon \cdot
\nabla \varepsilon
\end{equation}
where $\varepsilon = f + i \psi$, and $\nabla ^2$ and $\nabla$ are
the Laplacian and gradient in cylinderical coordinates.

Eq. (\ref{ernst}) is the Ernst equation \cite{ernst} of
general relativity. This form of the Bogomolny equations has
been used to find exact, nonsingular, multimonopole solutions
for the fields through the use of the B{\"a}cklund transformations
of Harrison \cite{harr}. Since the axial Bogomolny equations can
be written in the form of the Ernst equations, it should be
possible to use the known exact solutions of general relativity to
find exact solutions for SU(2) Yang-Mills-Higgs theory.
That this link between the general relativistic solutions and
their Yang-Mills counterparts has not been exploited before, can perhaps
be attributed to the singularities which exist in these solutions.
Here it is conjectured that these singularities might actually be a
desired feature in that they may provide a confinement mechanism for
non-Abelian gauge theories.
Previously \cite{sing1} \cite{sing2}, using a different approach
we have found exact Schwarzschild-like solutions for SU(2) and SU(N)
Yang-Mills-Higgs theories. Here we use the well known Kerr solution,
written in terms of variables of the Ernst equation, to give an equivalent
solution for the Yang-Mills theory. There are actually several other
exact, axially symmetric solutions in general relativity which could
be mapped over into Yang-Mills theory ({\it e.g.} the Tomimatsu-Sato
metric \cite{sato} and the NUT-Taub metric \cite{nut}). However the
Kerr metric is the simplest axially symmetric solution, and is of
physical interest since it gives the exterior gravitational field
for a central mass with angular momentum. However our axial
Yang-Mills solution apparently does not possess any angular momentum,
but rather seems to represent the non-Abelian field configuration
due to two concentric shells of SU(2) charge.
Thus although the Kerr-like solution is found using the
general relativistic solution, they appear to have some different
physical characteristics.

To find the Kerr solution from the Ernst equation one first introduces
the complex potential $\zeta$ such that
\begin{equation}
\label{sub1}
\varepsilon = f + i \psi = {\zeta - 1 \over \zeta +1}
\end{equation}
so that $f$ and $\psi$ are the real and imginary parts, respectively,
of $(\zeta -1)/(\zeta + 1)$. Substituting this expression for
$\varepsilon$ into, Eq. (\ref{ernst}), the Ernst equation becomes
\begin{equation}
\label{ernst1}
(\zeta {\bar \zeta} -1) \nabla ^2 \zeta = 2 {\bar \zeta} \nabla \zeta
\cdot \nabla \zeta
\end{equation}
where ${\bar \zeta}$ is the complex conjugate of $\zeta$.
For this form of the equations the Kerr solution is most easily found
using prolate spheroidal coordinates \cite{ernst}, which can be written
in terms of the cylinderical coordinates, $\rho$ and $z$, as
\begin{eqnarray}
\label{prosph}
x &=& {1 \over 2k} \left[ \sqrt{(z+k)^2 + \rho ^2} + \sqrt{(z-k)^2
+\rho ^2} \right] \nonumber \\
y &=& { 1\over 2k} \left[ \sqrt{(z+k)^2 + \rho ^2} - \sqrt{(z-k)^2
+ \rho ^2} \right]
\end{eqnarray}
where the inverse transformation is given by
\begin{eqnarray}
\label{prosph1}
\rho &=& k \sqrt{x ^2 -1} \sqrt{1 - y ^2} \nonumber \\
z &=&  kxy
\end{eqnarray}
where $k, p$ and $q$ are arbitrary constants. In these prolate
spheroidal coordinates the gradient and Laplacian become
\begin{eqnarray}
\nabla &=& {k \over \sqrt{x ^2 - y ^2}} \left[ {\hat{\bf x}}
\sqrt{x ^2 -1} {\partial \over \partial x} + {\hat{\bf y}}
\sqrt{1 - y ^2} {\partial \over \partial y} \right] \nonumber \\
\nabla ^2 &=& {k^2 \over x ^2 - y ^2} \left[ {\partial \over \partial
x} (x^2 -1) {\partial \over \partial x} + {\partial \over
\partial y}(1- y ^2) {\partial \over \partial y} \right]
\end{eqnarray}
where $\hat{\bf x}$ and $\hat{\bf y}$ are unit vectors.
Using these expressions it is easy to see that a solution to the second
form of the Ernst equation, Eq. (\ref{ernst1}) is
\begin{equation}
\label{soln}
\zeta = p x - i q y
\end{equation}
where the constants $p , q$ must satisfy the condition $p^2 + q^2 = 1$.
The above solution can be transformed into the standard form of the
Kerr solution by doing a transformation from the prolate spheroidal
coordinates to Boyer-Lindquist coordinates \cite{ernst}. The special
case when $q=0 , p =1$ and $\zeta = x$ gives (after a transformation to
Schwarzschild coordinates) the usual Schwarzschild metric for general
relativity. However using the solution $\zeta = x$ in the Yang-Mills
case to write down the expressions for the gauge and scalar fields,
we find that we do not recover our previous Schwarzschild-like solution
for SU(2), but obtain a different solution. The reason
for this lies in the fact that the ans{\"a}tze we used in each case
were different, and in the previous work we found our
solution directly from the Euler-Lagrange field equations, while here
we employed the Bogomolny formalism. A field configuration
that satisfies the Bogomolny equations will also satisfy the
Euler-Lagrange equations, but the reverse is not necessarily true.
Nevertheless $\zeta = x$ does give a solution to the
Yang-Mills-Higgs equations, which is a special case of the
general solution ({\it i.e.} $p,q \ne 0$) that we are considering.
In general relativity $p, q$ and $k$ are related
to the mass and angular momentum of the central mass which produces
the gravitational field. Here $p, q$ and $k$ will be related to the
shape of the axially symmetric
SU(2) charge configuration of our solution. In order
to find expressions for the fields $\phi _i , \eta _i , W_i$ it is first
necessary to determine the functions $f$ and $\psi$. From Eq. (\ref{sub1})
we find that
\begin{eqnarray}
\label{soln1}
f &=& {p^2 x^2 + q^2 y^2 - 1 \over (p x +1)^2 + q^2 y^2}
\nonumber \\
\psi &=& {-2 q y \over (p x +1)^2 + q^2 y ^2}
\end{eqnarray}
To get the gauge and scalar fields one simply inserts these expressions for
$f$ and $\psi$ into Eq. (\ref{gfields}). This is
a straightforward but tedious procedure which yields very complicated
expressions for the fields. The explicit expressions for the fields are
\begin{eqnarray}
\label{gfields1}
\phi _1 &=& - W_1 = {-2 q \Big[x (1-y^2)\big((px +1)^2 -q^2
y ^2 \big) -2p y^2 (p x +1) (x ^2 -1) \Big] \over
k [ (p x +1) ^2 + q^2 y ^2 ] (p^2 x ^2 +q^2 y ^2
-1) (x ^2 - y ^2) } \nonumber \\
\space \space \nonumber \\
\eta _1 &=& \rho W_2 = {-2 q y \Big[(p x  +1)^2 -q^2
y ^2 + 2p x (p x +1) \Big] (x ^2 -1)
(1- y ^2) \over [ (p x +1)^2
+ q^2 y ^2 ] (x ^2 - y ^2) (p^2 x^2 + q^2 y^2
-1)} \nonumber \\
\space \space \nonumber \\
\phi _2 &=& {-2 xy [ p^2(x ^2 -1) + q^2(1 - y ^2)]
\over k (x^2 - y^2)(p^2 x ^2 + q^2 y^2 -1) } +
{2 y [ p(p x +1)(x ^2 -1) + q^2 x (1- y ^2) ]
\over k (x^2  - y ^2) [ (px +1)^2 + q^2 y ^2 ] }
\nonumber \\
\space \space \nonumber \\
\eta _2 &=& {2 (x^2 -1) (1- y^2) (p^2 x^2 - q^2 y^2)
\over (x^2 - y^2)(p^2 x^2 + q^2 y^2 -1) } -
{2 (x^2 -1)(1- y^2) [p x (p x +1) - q^2 y ^2]
\over (x ^2  - y ^2) [(p x +1)^2 + q^2 y ^2]}
\end{eqnarray}
where the partial derivatives like, $\partial \psi / \partial z$,
were determined using the chain rule
({\it e. g.} $\partial _x \psi \partial _{z} x$) and Eq.
(\ref{prosph}). Notice that in general relativity the physical
quantities one usually deals with are the components of the metric
tensor. Here the physical quantities are the gauge fields which
correspond to the Christoffel symbols in general relativity.
This partly explains the complexity of the expressions in Eq.
(\ref{gfields1}), since even in Boyer-Lindquist coordinates, the
Christoffel coefficients for the Kerr metric are somewhat
involved. If one wanted to have the expressions for the fields
in terms of the original coordinates, it would be necessary to use
Eq. (\ref{prosph}) to replace $x, y$ with $\rho , z$, making
an already complicated expression even more intractible. However by
looking at certain aspects of the expressions of the fields one can
still make some interesting comments about this solution.

One feature that can be looked for are regions where the fields
become singular. In analogy with Maxwell's equations, where
singularities in the electromagnetic field indicate the presence
of electric charge, we interpret these singularities as the
location of color charge. The shape of these SU(2) charge
distributions is much more involved than in electromagnetism.
All of the fields from Eq. (\ref{gfields1}) have three
similiar terms in their denominators, so the fields
can be made to approach infinity if any one of the three
factors goes to zero. First, by setting
$p^2 x^2 + q^2 y^2 -1 = 0$ it can be seen that
all the fields become infinite. This factor is common
to all the fields, since from Eq. (\ref{gfields}) they all
have a factor of $f ^{-1}$. In cylinderical coordinates
this condition becomes
\begin{equation}
\label{singular}
\vert pq \vert (z^2 + \rho ^2 - k^2) = \pm
\vert q^2 - p^2 \vert \rho k
\end{equation}
By setting $q = cos \theta$ and $p = sin \theta$ we replace the
two parameters $p, q$ with one parameter, and Eq. (\ref{singular})
becomes
\begin{equation}
\label{singular1}
z^2 + \rho ^2 - k^2 = \pm 2 \vert cot (2 \theta ) \vert \rho k
\end{equation}
Solving the
above condition for $z$ as a function of $\rho$ allows one
to take a vertical slice through the two axially symmetric
surfaces defined by Eq. (\ref{singular1}). One needs only
to look in the range, $0^o \le \theta \le 45 ^o$ to cover
all the possibilities. What one finds are two concentric
surfaces which which touch each other on the $z$-axis at
$\pm k$. The outer surface is given by the positive solution
to Eq. (\ref{singular1}). It has a toroidal shape, without
the central hole of a normal torus. The inner surface is given
by the negative solution, and has an ellipsoidal shape which runs
along the $z$-axis. In the special case when $\theta = 45 ^o$
({\it i.e.} $p=q$) the two surfaces merge into a single
sphere with a radius of $k$. Second,
the fields can become infinite if $x ^2 - y ^2 = 0$.
However this condition gives two points ($\rho = 0$, $z = \pm k$)
which are already included in the first condition of Eq.
(\ref{singular}). Finally some of the fields become singular when
$y = 0$ and $px + 1 =0$ (since $x$ is positive definite this
condition only has a solution when $p < 0$). The condition
$y =0$ implies $z=0$ so that the singularity resides in
the plane perpendiuclar to the $z$-axis,
and then $p x + 1 = 0$ gives $\rho = k \vert q \vert
/ \vert p \vert = k \vert cot \theta \vert$. This corresponds
to a ring singularity of radius $k \vert cot \theta \vert$
centered at the origin in the plane perpendicular to
the $z$-axis. However this singularity, produced by the
condition $y=0$ and $px +1 =0$, duplicates that from the
condition of Eq. (\ref{singular1}).
Thus the singularity produced by the third term in the
denominators of the gauge fields does not produce
an independent singularity. The geometrical
structure of the singular surfaces
is different from that of the similiar Schwarzschild-like
solution. For the Schwarzschild solution we obtained a
spherical shell singularity surrounding a point singularity
at the origin. In the present case we find concentric
toroidal and ellipsoidal surfaces on which the gauge and scalar
fields become infinite, and with no apparent singularity
in the interior of these surfaces.

The connection of both the Schwarzschild-like solution and the
Kerr-like solution with confinement can be seen
both classically and quantum mechanically. First, from a classical
point of view, the non-Abelian fields become arbitrarily large
as one approaches the surfaces defined by Eq. (\ref{singular}).
Any particle which carries an SU(2) charge would either be
strongly attracted or strongly repelled as it approached
these surfaces of Eq. (\ref{singular}), depending on the relative
``sign'' of the SU(2) charge of the particle to that of the surface.
If the force between the particle and surface were replusive, the
particle would never be able to enter the interior region. If the
force were attractive then once the particle entered the interior
region it would no longer be able to leave, thus becoming
permanently confined. This behaviour in the attractive case is
similiar to the behaviour of particles in the vicinity of a
general relativistic black hole. The replusive case is
different from that of general relativity since Yang-Mills
theories are vector theories, which lead to both attractive
and replusive forces, whereas general relativity is a tensor
theory which leads only to attractive forces.
Quantum mechanically the above configuration is similiar in
character to some phenomenological bag models of hadron
dynamics, where QCD bound state are modeled as free particles
({\it e.g.} quarks) inside a spherical hard wall potential.
In the case of the Schwarzschild-like solution it has been
shown \cite{lunev} that if a scalar
or spinor particle with SU(2) charge
is placed in the potential of the gauge fields it will remain
confined to the region inside the surfaces of infinite
field strength. Our analytic Schwarzschild-like
solution \cite{sing1} is slightly  different from
those of Ref. \cite{lunev} in that our solution is singular
at the origin. However since the surface singularity is a common
feature of both solutions we expect that our solution will
also permanently confine any color charged particle.

The exact expressions for the
energy and angular momentum of the Kerr-like solution
presented here are rather complicated due to the involved
nature of the scalar and gauge fields (see Eq.
(\ref{gfields1})). Still some interesting general conclusions
can be made about these quantities. To find the energy and
angular momentum in the fields it is necessary to calculate
the energy-momentum tensor of the Lagrangian of Eq. (\ref{lagrange})
\begin{equation}
\label{emt}
T^{\mu \nu} = {2 \over \sqrt{-g}} {\partial ({\cal L} \sqrt{-g})
\over \partial g_{\mu \nu}} = F^{\mu \rho a} F_{\rho} ^{\nu a}
+ D^{\mu} \phi ^a D^{\nu} \phi ^a + g ^{\mu \nu} {\cal L}
\end{equation}
The energy of the field configuration is then
\begin{equation}
\label{energy}
E = \int d^3 x T^{00} = \int d^3 x
\left[ {1 \over 4} F_{ij} ^a F^{aij} + {1 \over 2}
F_{0i}^a F^{a0i} + {A^2 \over 2} D_i \phi ^a D^i \phi ^a +
{A^2 \over 2} D_0 \phi ^a D^0 \phi ^a \right]
\end{equation}
This is equivalent to the Hamiltonian except the signs are all
positive. Now using $D^0 \phi ^a = 0$, $F_{0i} ^a = C (D_i \phi ^a)$
and $F_{ij} ^a = \sqrt{A^2 - C^2} \epsilon_{ijk} D^k \phi ^a$ the
energy becomes
\begin{equation}
\label{energy1}
E = A^2 \int d^3 x D_i \phi ^a D^i \phi ^a
\end{equation}
The constant $A$ is the multiplicative factor that we put in front
of the scalar fields in order that we could easily examine the
case when there were no scalar fields by taking $A = 0$.
{}From Eq. (\ref{energy1}) it can be seen that for this special case
the energy in the fields of the Kerr-like solution is zero.
In addition when $A=0$ either the time components
or the space components of the gauge fields
are pure imaginary. These results were the same for the
Schwarzschild Yang-Mills solution, and we believe that this calls
into question the physical relevance of the pure gauge case even
though it is mathematically a solution. Taking this view then requires
the presence of a scalar field in order to get a physically
``reasonable'' solution. The angular momentum of the field
configuration is given by
\begin{equation}
\label{angmom}
L = \int d^3 x \epsilon_{ijk} x^j T^{0k}
\end{equation}
Using the expression for $T^{\mu \nu}$ from Eq.(\ref{emt}),
the condition $F_{0i}^a = C (D_i \phi ^a)$ and the Bogomolny
field equations, Eq. (\ref{bogo}), we find
\begin{equation}
T^{0k} = \epsilon ^{klm} C \sqrt{A^2 - C^2} (D_l \phi ^a)
(D_m \phi ^a)
\end{equation}
The antisymmetry  of $\epsilon^{klm}$ makes $T^{0k} = 0$,
so that there is no angular momentum in this non-Abelian
field configuration. This shows that while our Yang-Mills
solution is similiar in many ways to its general relativistic
counterpart, there are some important distinctions. Part of
the reason for this stems from the fact that the symmetries of
general relativity are space-time symmetries, while those
of Yang-Mills theories are internal Lie symmetries. These
differences showed up even in the Yang-Mills Schwarzschild-like
solution where the sphere singularity was a true singularity,
while for general relativity the event
horizon is a coordinate singularity, as can be seen by looking
at the Schwarzschild solution in Kruskal coordinates.

\section{Discussion and Conclusion} Extending our previous
work on the Schwarzschild-like solution for Yang-Mills-Higgs
theories, we have written down the Yang-Mills equivalent
of the Kerr solution. By writing the Yang-Mills field equations
in the form of the Ernst equation \cite{forg} of general
relativity \cite{ernst} it is straightforward to use any
known axially symmetric solution of general relativity to
write down similiar solutions in terms of the
non-Abelian gauge fields.
One disadvantage of these general relativistic inspired
solutions is that they contain singularities in the fields,
which lead to infinite field energies at the classical level.
These singularities are of the same character as the singularities
which are found in other classical field theory solutions such
as the singularity at the origin in the normal Schwarzschild
solution, the point singularity in the Wu-Yang solution for SU(2)
\cite{wu}, and the singularity at $r=0$ in the Coulomb
potential in electromagnetism.
Since our solutions are classical field theory
solutions it may be conjectured, as is the case with general
relativity, that a proper quantum treatment of the problem
might modify these singularities. Fortunately,
unlike  the case of general relativity, there
do exist methods for quantizing such classical solutions
\cite{lee} \cite{woo}. In one sense however, these
singularities (particularly the surface singularities) are a
desirable feature in that they may yield a possible confinement
mechanism for non-Abelian gauge theories, which would be
analogous to the confinement mechanism of general relativistic
black holes. Quantum
mechanically it can be shown that these solutions
lead to a simple form of confinement, by placing a particle
with color charge into the potentials of our solutions and
solving the relevant quantum problem. This was outlined in
Ref. \cite{lunev} where color charged test particles were
placed in a Yang-Mills field configuration similiar in some
respects to ours.
It was found that the wavefunction of the test particle
was confined to the region around the origin, which is
in accord with our initial findings using our Schwarzschild-like
solution. While this quantum mechanical analysis is suggestive,
it does have shortcomings. The chief one being that it ignores
the interaction between the color field of the test particle
and the field configuration in which it is placed. Since the
field equations of Yang-Mills theories are non-linear one is not
justified in using superposition, and the color field of the
test particle could significantly alter the color field of the
solution. This is similiar to the problem in general relativity
of the interaction of two black holes, which must be handled
numerically.  Nevertheless one could argue that two color charged
particles could form a bound state in such a way that they move
in an average field configuration which is given approximately
by our solutions.

The very direct link, via the Ernst equations, between general
relativity and Yang-Mills-Higgs theories can be used to map
over any of the known axially symmetric solutions from general
relativity into non-Abelian gauge theories. In this paper we
examined in detail only the Kerr solution in the hope of finding
a field configuration, which contained an internal angular
momentum, so that quantizing this angular momentum to $\hbar /2$
one would be able to have a fermion-like object from an initial
theory with only gauge and scalar fields. Explicitly carrying
out the calculation of the field angular momentum showed that
even though the Kerr-like solution was axially symmetric, it did
not carry any angular momentum in its fields.
This shows that not all the
features of the general relativistic solution carry over into
Yang-Mills theory. So far we have used this parallel between
the two theories to find solutions for Yang-Mills theories from
the known  solutions of general relativity. An interesting
exercise might be to see if some of the the exact solutions of
Yang-Mills theory ({\it e.g.} the Prasad-Sommerfield solution
or the multimonopole solutions \cite{forg1}) could be used to
give unknown exact solutions in general relativity.

\section{Acknowledgements} The author wishes to acknowledge the
help and suggestions of Donald Singleton and Heather O'Neill.

\end{document}